\documentstyle[11pt,newpasp,twoside,epsf]{article}
\def\msun{{M_\odot}}

\def\edcomment#1{\iffalse\marginpar{\raggedright\sl#1\/}\else\relax\fi}
\reversemarginpar

\begin{document}
\title{What X-rays tell us about dark matter halos}
 \author{Y.P. Jing}
\affil{Shanghai Astronomical Observatory, the Partner Group of 
MPI f\"ur Astrophysik, Nandan Road 80, Shanghai 200030, China}

\begin{abstract}
We present a detailed non-spherical modeling of dark matter halos on
the basis of a careful analysis of state-of-the art N-body
simulations.  The fitting formula presented here form a complete and
accurate description of the triaxial density profiles of halos in Cold
Dark Matter(CDM) models. This modeling allows us to quantitatively discuss
implications for shape observations of galactic and cluster halos. The
predictions of the concordance $\lambda$CDM model are confronted with the
shape observations from the Milky Way to X-ray clusters.
\end{abstract} 
%
%%%%%%%%%%%%%%%%%%%%%%%%%%%%%%%%%%%%%%%%%%%%%%%%%%%%%%%%%%%
\section{Introduction}
%%%%%%%%%%%%%%%%%%%%%%%%%%%%%%%%%%%%%%%%%%%%%%%%%%%%%%%%%%%

\begin{figure}
\plottwo{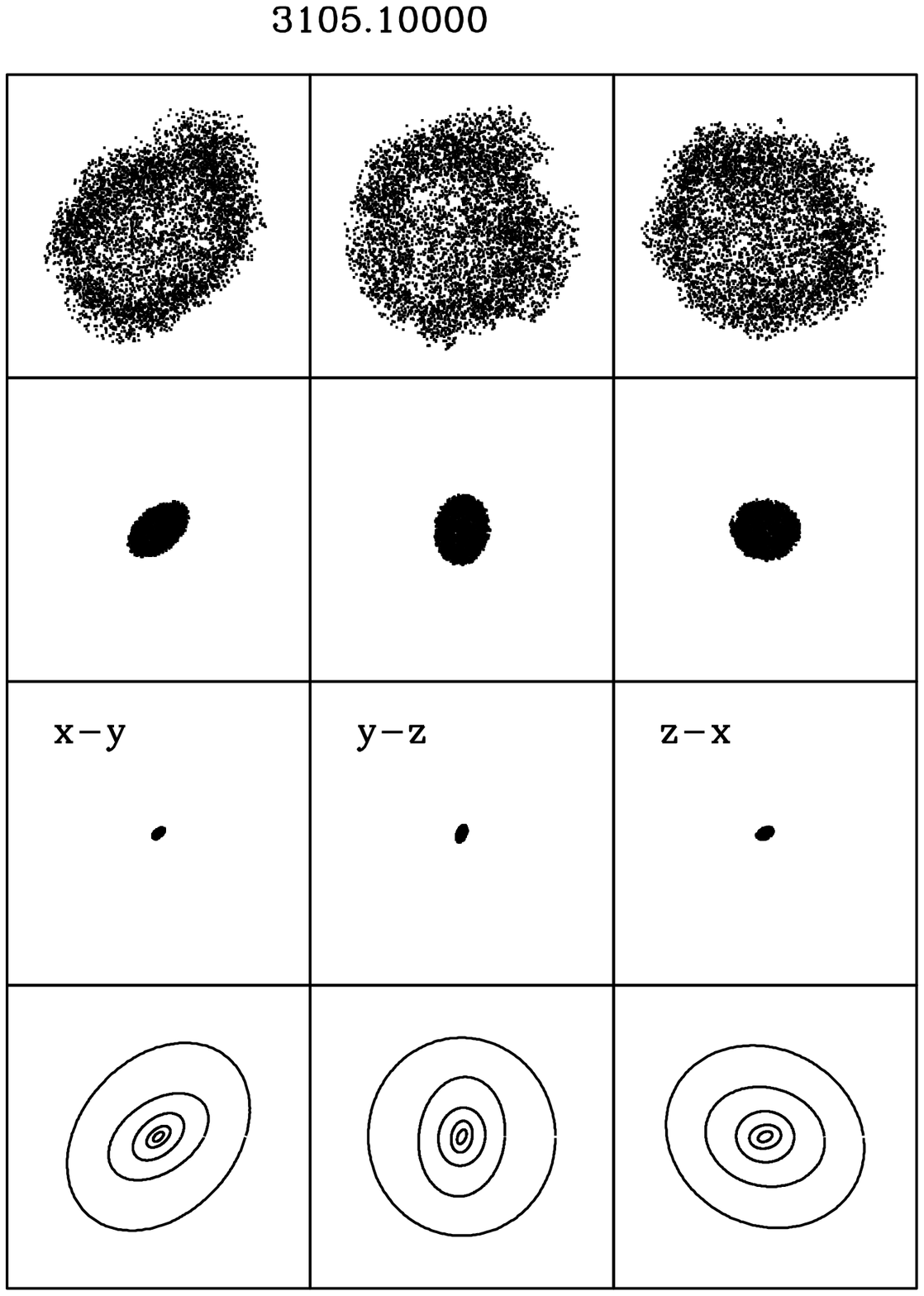}{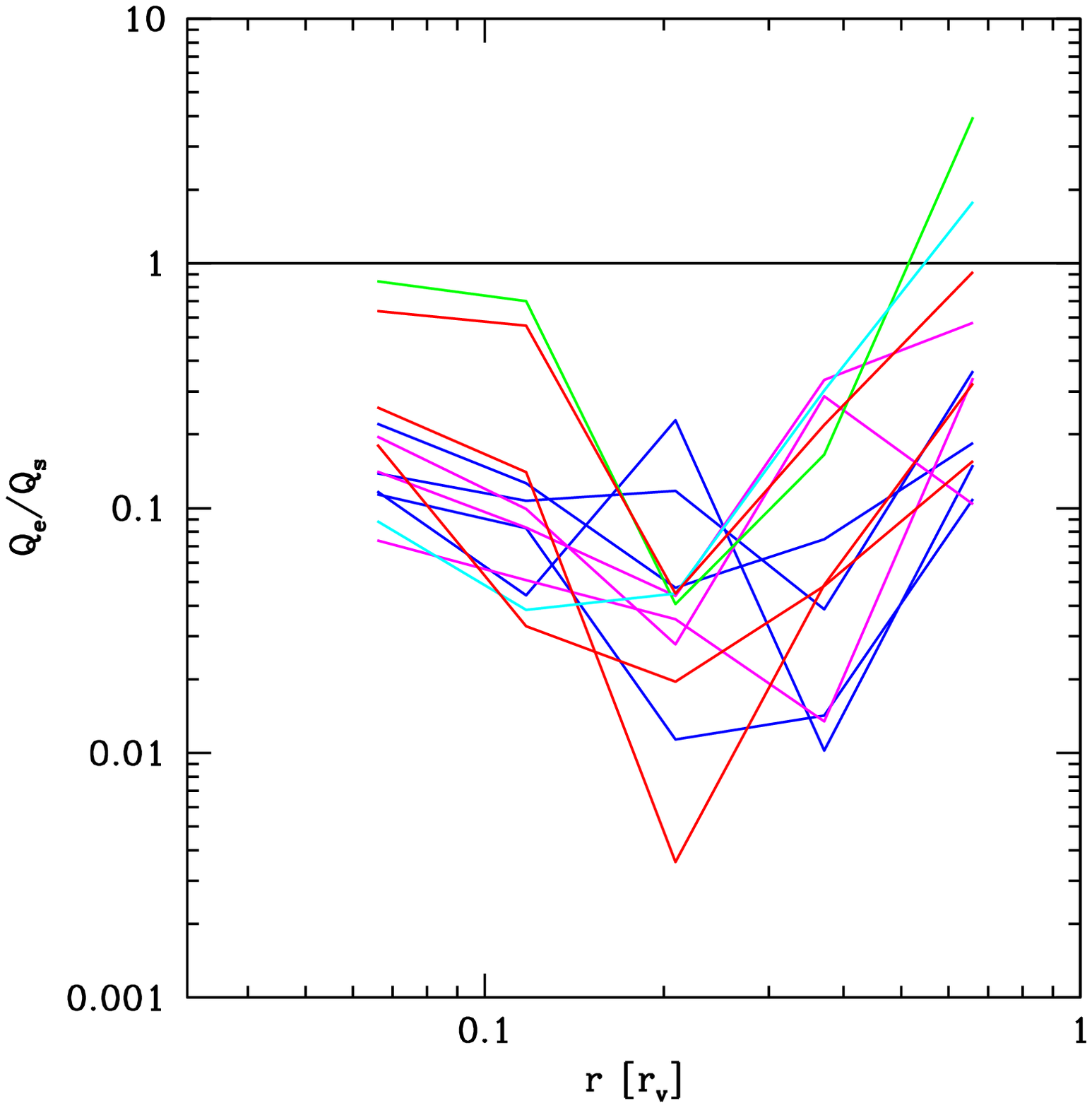}
\caption{{\it Left panel} - A typical example of the projected
iso-density surfaces within a dark halo. The bottom panels show the
triaxial fits to five isodensity surfaces projected on those
planes. {\it Right Panel} - Ratio of the quadrupole moments defined in
the triaxial model ($Q_e$) and in the spherical model ($Q_s$) for five
shells at radii from $0.05r_{\rm vir}$ to $0.65r_{\rm vir}$. The
figure clearly shows that the triaxial model describes the density
profiles of halos more accurately than the conventional spherical
model. From Jing \& Suto (2002).  }

\end{figure}

It is well known that dark matter halos are not spherical as predicted
by current theories of structure formation (V. Springel, this volume,
for a review). The shape of halos depends on the nature of dark matter
and cosmological parameters in theory, and can be measured through a
wide range of observations from the kinematics of satellites and
debris in galaxies to the hot gas distribution of galaxy clusters. In
this talk, I will first present an accurate model for the shape of
dark halos in Cold Dark Matter (CDM) dominated models, and then
discuss its implications for and confrontations with observed shapes
of galactic and cluster halos.

%%%%%%%%%%%%%%%%%%%%%%%%%%%%%%%%%%%%%%%%%%%%%%%%%%%%%%%%%%%%%%%%%%%%%%%%%%%
\section{Modeling the non-spherical density profiles of dark matter halos}
%%%%%%%%%%%%%%%%%%%%%%%%%%%%%%%%%%%%%%%%%%%%%%%%%%%%%%%%%%%%%%%%%%%%%%%%%%%
We use two sets of state-of-the art simulations for the current
purpose. The first is a set of cosmological N-body simulations
with $N=512^3$ particles in a $100h^{-1}$Mpc box (Jing \& Suto 2002),
and the other is a set of high-resolution halo simulation runs (Jing
\& Suto 2000).

To model the shape of dark matter halos, we first find the iso-density
surfaces.  This begins with a computation for the local density at
each particle's position.  We adopt the smoothing kernel widely
employed in the Smoothed Particle Hydrodynamics (SPH, Hernquist \&
Katz 1989).  Thirty two nearest neighbor particles are used to compute
the local density.  The left panel of Figure~1 shows a typical example
of the halo isodensity surfaces.  This plot clearly suggests that the
isodensity surfaces can be approximated as triaxial ellipsoids.

The isodensity ellipsoids at different radii are well aligned, and the
axial ratios of the ellipsoids are nearly constant. These facts
suggest that the internal density distribution within a halo can be
approximated by a sequence of the concentric ellipsoids of a constant
axis ratio. To show this to be an improved description over the
conventional spherical description, we compute the quadrupole of the
particle distribution within a spherical shell ($Q_s$) or an
ellipsoidal shell ($Q_e$). If the spherical (triaxial) model is exact,
$Q_s$ $(Q_e)$ vanishes, thus a more accurate halo density profile
should yield a smaller quadrupole.  The ratio of these two quantities
is plotted in the right panel of Figure 1. It clearly shows that the
triaxial model works much better than the conventional spherical
model.

We found that the matter density within the isodensity surfaces
changes with the major axis $R$ in a way similar to that of the
spherical model. The density profile can be approximately described by
the NFW-like profiles, though the concentration parameter is different
from the spherical model. In Jing \& Suto (2002), we have given a
recipe, based on our analysis of the simulations, for predicting the
triaxial density profile for a halo in a general CDM cosmogony if the
halo mass and halo shape (i.e. the axial ratios) are known.

The mass distribution of CDM halos is well described by the
Press-Schechter formula (or its modified form) for theories of
formation. Thus, in order to predict the density profiles for a sample
of CDM halos, we need to quantify the distribution function
$p(a/c,b/c)$ of the axial ratios $a/c$ and $b/c$ at a given mass,
where $a$, $b$, and $c$ are the minor, middle, and major axes
respectively.  The function can be written as
\begin{eqnarray} 
\label{eq:pcon} 
p(a/c,b/c)d(a/c)d(b/c) &=& p(a/c)d(a/c) ~ p(b/c|a/c)d(b/c) \cr
&=& p(a/c)d(a/c) ~ p(a/b|a/c)d(a/b)\,.
\end{eqnarray} 
The probability function $p(a/c)$ and the conditional probability $
p(a/b|a/c)$ have been measured for halos in our SCDM and $\lambda$CDM
simulations at different epochs. An example is given in Figure 2.
Jing \& Suto (2002) have found accurate universal functions to
describe these two probability functions. The shapes depend both on
the halo mass (through the ratio of the halo mass to the non-linear
mass $M_*$) and on the cosmological parameters, and the distribution
of the shapes is very broad. This means that in order to test the
shape prediction of CDM models definitely, one generally needs a large
sample of objects, e.g. galaxies or clusters of galaxies. As a
reference, it would be interesting to have a look at typical values of
halo shapes. For the concordance $\lambda$CDM model of a fluctuation
amplitude $\sigma_8=0.9$, the typical values of the axial ratios $a/c$
and $b/c$ are 0.7 and 0.8 for Milky Way halos and 0.45 and 0.65 for
clusters of galaxies. The galactic halos at redshift $z=0.5$ are
expected to have $a/c=0.5$ and $b/c=0.7$, much more eccentric than
their counterpart at $z=0$.

\begin{figure}{t}
\plottwo{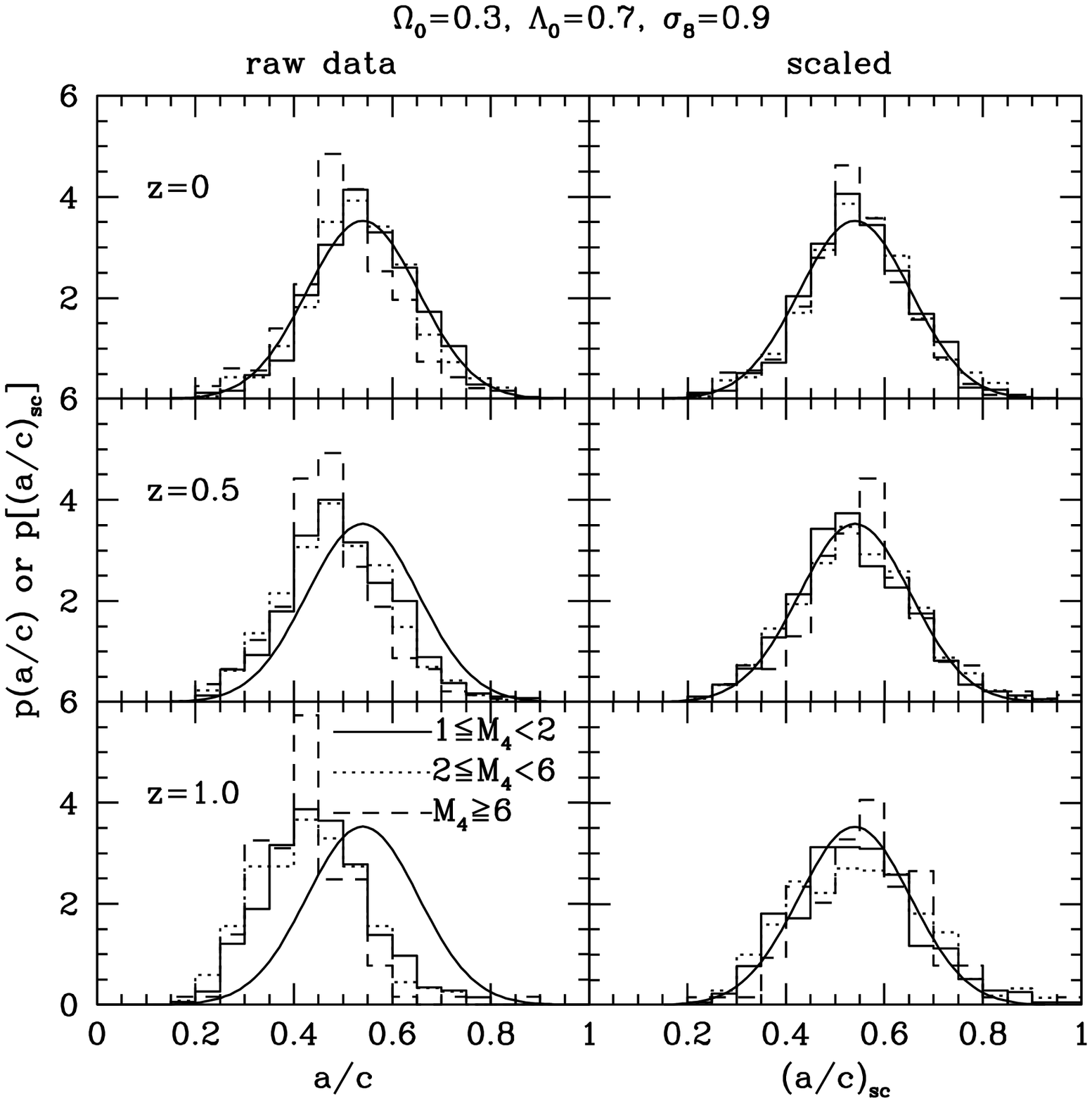}{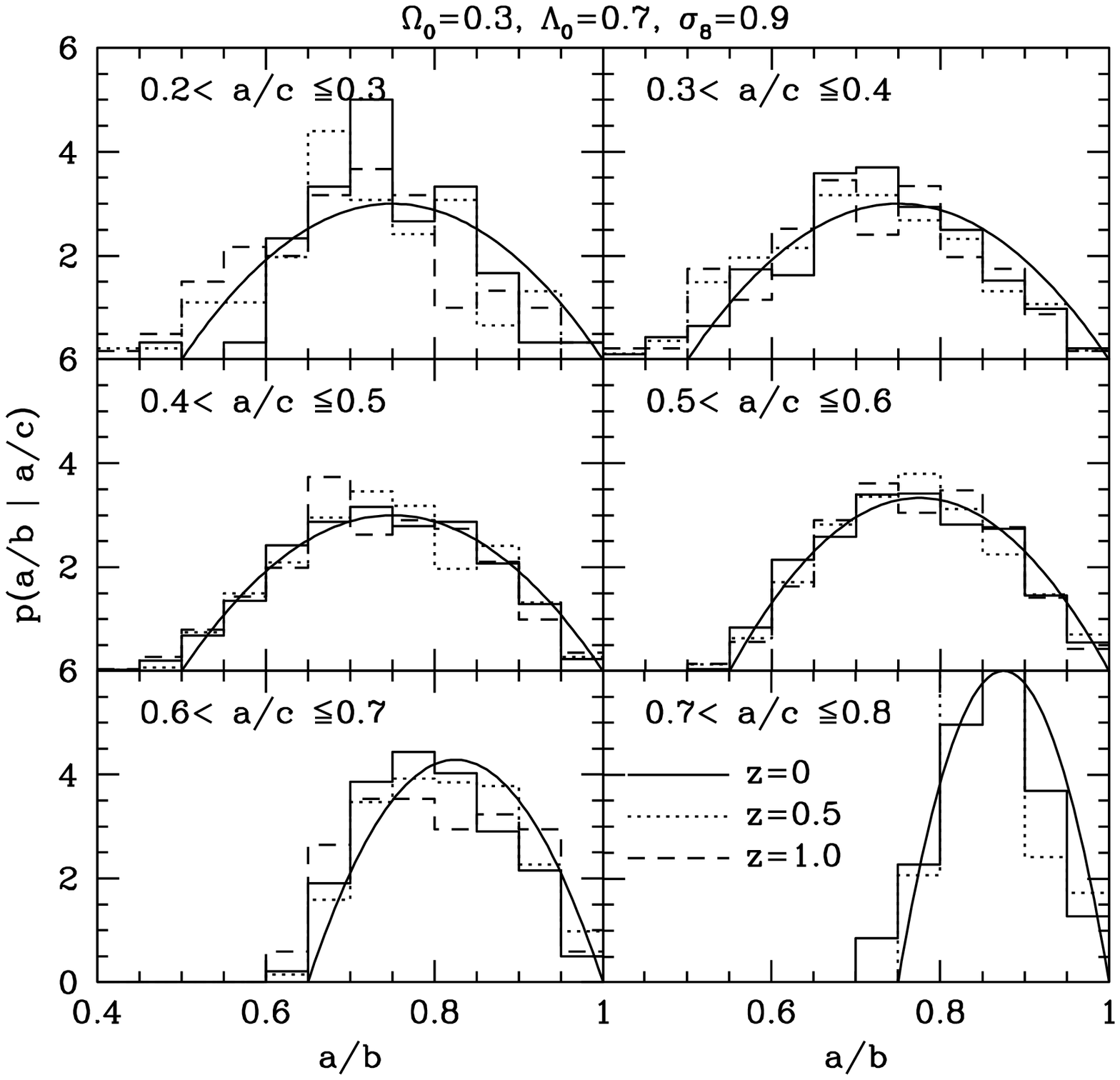} 
\caption{{\bf Left Figure}- The distribution of the axis ratio $a/c$
of the halos in the cosmological simulations of the $\lambda$CDM model
before ({\it left panels}) and after ({\it right panels}) the scaling
is applied (see Jing \& Suto, 2002).  Solid, dotted and dashed
histograms indicate the results for halos that have $M_4 \times 10^4$
particles within the virial radius. {\bf Right Figure} - The
conditional distribution of the axis ratio $a/b$ of the halos in the
cosmological simulations of the $\lambda$CDM model for a given range
of $a/c$.  The smooth solid curves are the fitting formulas. From Jing
\& Suto (2002).}

\end{figure}

\section{Comparison with the observations}
From the kinematics within galaxies or from the X-ray observations of
clusters, one may measure the potential of their dark halos. For the
triaxial model in the last section, Lee \& Suto (2003) presented a
calculation for the distribution of iso-potential surfaces. Using the
eccentricity $e = (1-a^2/c^2)^{1/2}$, the ratio of the halo potential
eccentricity to the halo density eccentricity at radius $r$ is a
function of $r/R_0$ (Lee \& Suto 2003; the left panel of Figure 3),
where $R_0$ is a scale radius of the major axis in the generalized NFW
form (Jing \& Suto 2002). The potential becomes rounder at an outer
radius, because the halo mass is concentrated in the central
region. For a galactic halo of Milky Way mass, the axial ratio $a/c$
of the potential is 0.85 at the central region and is larger than 0.9
at the virial radius. These values are well consistent with the
potential shape measured for the Milky Way (Sackett, P., Sellwood, J.,
this volume).

\begin{figure}{t}
\plottwo{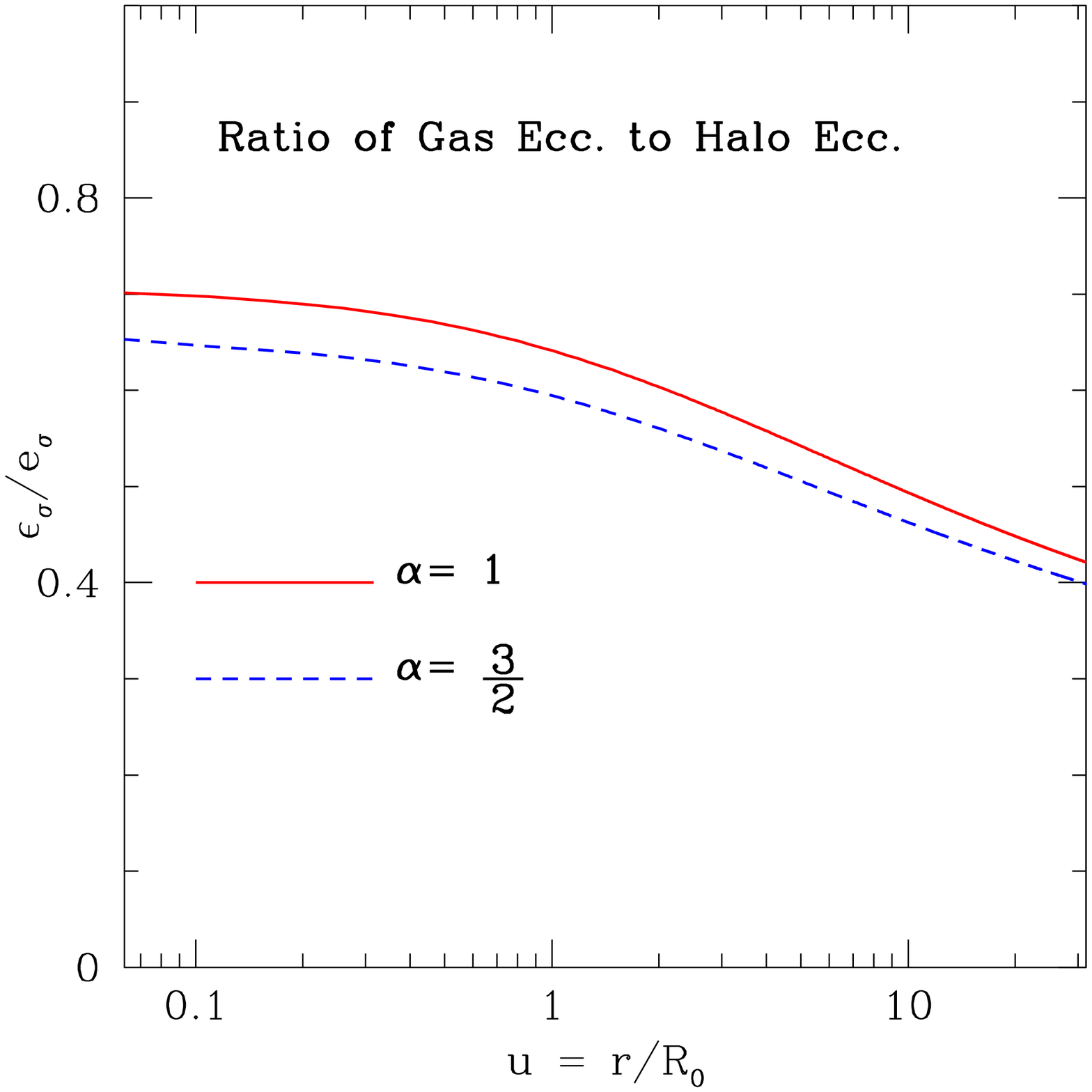}{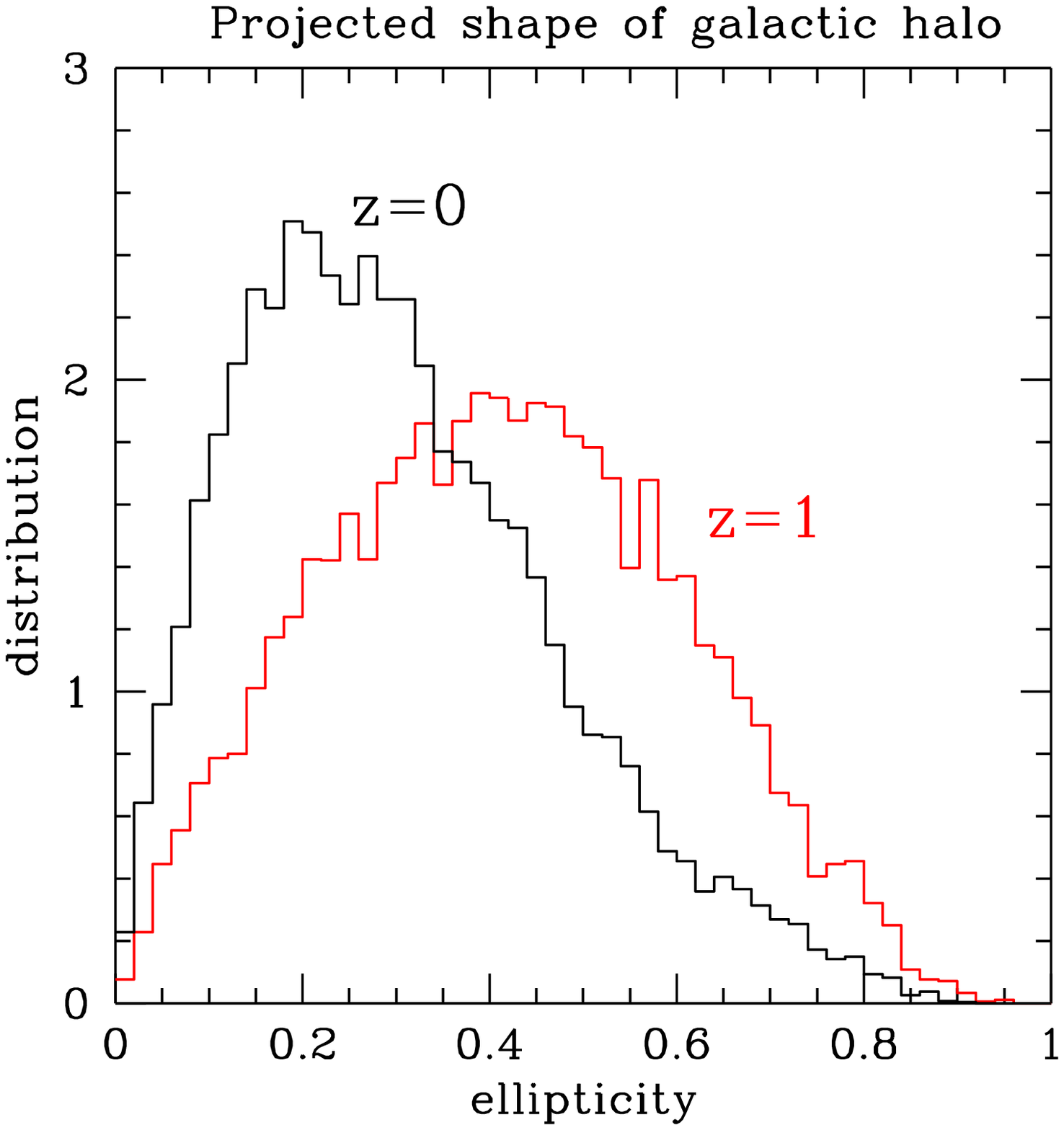} 
\caption{{\bf Left Figure}- The ratio of the eccentricity of the
iso-potential surface to that of the halo iso-density surface from the
perturbative result for the NFW profile ($\alpha = 1$) and for the
Moore et al. like profile ($\alpha = 3/2$, dashed line). From Lee \&
Suto (2003). {\bf Right Figure} - The ellipticity distribution of
projected galactic halos at $z=0$ and $z=1$ in the concordance
$\lambda$CDM model.}
\end{figure}

X-ray morphology of clusters can be computed from the potential under
the hydrostatic equilibrium assumption, or can be measured directly
from hydro/N-body simulations. Flores et al. (in preparation) have
carried out a comparison of the cluster ellipticity between the
concordance $\lambda$CDM model and an Einstein X-ray cluster
sample. The distribution of the model clusters is completely
consistent with that of the observed clusters, though larger samples
both of simulation clusters and of observed clusters are badly needed
for an accurate assessment. Flores et al. also noted that the shape
distribution found in their work is well consistent with Jing \& Suto
(2002).  In a recent work of Plionis (2002), a significant evolution
is found of the cluster X-ray shapes, as higher-redshift clusters are
more elongated than the local clusters.  Floor et al. (2003) found it
difficult to explain this ellipticity evolution within the concordance
$\lambda$CDM model. I would like to point out, however, the comparison
between the model and the observation carried out by Floor et
al. might not be proper, because they used individual cluster
simulations in their study. The ``clusters'' they identified at high
redshift must be smaller than the clusters at $z=0$, while in the
observation of Plionis (2002) clusters at high redshift are much more
massive than those at $z=0$. With the triaxial model of Jing \& Suto
(2002), it is not difficult to explain the redshift dependence of the
cluster ellipticity found by Plionis (2002), at least qualitatively,
because the high redshift clusters in the observation are more
massive.

It is promising to probe the halo shapes directly with the weak
lensing effect. The weak lensing measures the projected mass
distribution (Schneider, P., this volume; Gavazzi et al. 2003).  We
have measured the shapes for the projected halo mass distribution in
the simulations. The right panel of Figure 3 shows the distribution of
the projected ellipticity for galactic halos (mass about
$10^{12}\msun$) in our simulations. The halos have a mean ellipticity
$\epsilon=0.31$ or axial ratio $a/b=0.72$ at redshift $z=0$, and
$\epsilon=0.42$ or $a/b=0.64$ at $z=1$. Recently, Hoekstra et
al. (2003) have tentatively measured the shape of galactic halos
through weak galaxy-galaxy lensing, and found $a/b=0.66$ for galactic
halos at $z\approx 0.5$. Their results are in good agreement with our
theoretical predictions, though it is important to note that they used
a weighting procedure to measure the halo shapes that must be properly
modeled in future comparisons between the observation and the
theoretical predictions.

In addition, we have compared the shape distribution of rich clusters
in the Standard Cold Dark Matter (SCDM; $\Omega_0=1$ and
$\lambda_0=0$) and in the concordance CDM ($\Omega_0=0.3$ and
$\lambda_0=0.7$), and found their difference is not dramatic. One
needs a large sample of local massive clusters (say a few thousand) to
distinguish these two models definitely. This conclusion is in good
agreement with an early study of Jing et al. (1995) on this subject.

\section{Conclusions}
We have presented a triaxial modeling of the dark matter halo
density profiles on the basis of the combined analysis of
the high-resolution halo simulations (12 halos with $N\sim 10^6$
particles within their virial radius) and the large cosmological
simulations (5 realizations with $N=512^3$ particles in a $100h^{-1}$Mpc
boxsize). In particular, we found that the universal density profile
discovered by NFW in the spherical model can be also generalized to our
triaxial model description. Our triaxial density profile is specified by
the concentration parameter $c_e$ and the scaling radius $R_0$ (or the
{\it virial} radius $R_e$ in the triaxial modeling) as well as the axis
ratios $a/c$ and $a/b$.

We have obtained accurate fitting formulae for those parameters which
are of practical importance in exploring the theoretical and
observational consequences of our triaxial model. Because the page
limit of the proceedings, we could not have typed in these
formulae. We refer the interested reader to the journal paper of Jing
\& Suto (2002) for the fitting formulae. As Springel summarized in his
talk (this volume), the shape distribution of dark halos is now well
determined in CDM models.

The model predictions are compared with various observations of halo
shapes around galaxies or clusters of galaxies. While a good agreement
is found between the concordance model prediction and the available
observations, a large sample of galactic and cluster halos is needed
to definitely test the prediction of the concordance model or to
distinguish among theories of different dark matter, because the shape
distribution of halos are generally very broad. It is also important
to point out that iso-potential surfaces are much rounder than the
iso-density surfaces, especially at halo outer part $r\approx
r_{vir}$.

%%%%%%%%%%%%%%%%%%%%%%%%%%%%%%%%%%%%%%%%%%%%%%%%%%%%%%%%%%%%%%%
\acknowledgments
%%%%%%%%%%%%%%%%%%%%%%%%%%%%%%%%%%%%%%%%%%%%%%%%%%%%%%%%%%%%%%%

I would like to thank the collaborator Yasushi Suto for his important
contribution to the work, and Joel Primack for communicating their
result on the shape distribution of x-ray clusters before publication.
The work was supported in part by by NKBRSF (G19990754) and by
National Science Foiundation of China.

\end{document}